\documentclass[final,5p,times,twocolumn]{elsarticle}

\usepackage{graphics}

\usepackage{amssymb}

\usepackage{lineno}

\usepackage{dcolumn}
\usepackage{bm}
\usepackage{hyperref}

\journal{Journal of Magnetism and Magnetic Materials}

\begin{document}

\begin{frontmatter}



\title{Micromagnetics of shape anisotropy based permanent magnets}


\author[fhstp]{Simon Bance}
\ead{s.g.bance@gmail.com}
\ead[url]{http://academic.bancey.com}
\author[fhstp]{Johann Fischbacher}
\author[fhstp]{Thomas Schrefl}
\author[siemens]{Inga Zins}
\author[siemens]{Gotthard Rieger}
\author[siemens]{Caroline Cassignol}

\address[fhstp]{St. P\"olten University of Applied Sciences, Matthias Corvinus Str. 15, 3100 St. P\"olten, Austria }
\address[siemens]{Siemens AG, Corporate Technology, Otto-Hahn-Ring 6, 81739 Munich, Germany}

\begin{abstract}
In the search for rare-earth free permanent magnets, various ideas related to shape anisotropy are being pursued. In this work we assess the limits of shape contributions to the reversal stability using micromagnetic simulations.  
In a first series of tests we altered the aspect ratio of single phase prolate spheroids from 1 to 16. Starting with a sphere of radius $4.3$ times the exchange length $ L_{\mathrm{ex}}$ we kept the total magnetic volume constant as the aspect ratio was modified. 
For a ferromagnet with zero magnetocrystalline anisotropy the maximum coercive field reached up to $0.5$ times the magnetization $M_{\mathrm{s}}$. 
Therefore, in materials with moderate uniaxial magnetocrystalline anisotropy, the addition of shape anisotropy could even double the coercive field.  
Interestingly due to non-uniform magnetization reversal there is no significant increase of the coercive field for an aspect ratio greater than 5. 
A similar limit of the maximum aspect ratio was observed in cylinders.  The coercive field depends on the wire diameter. By decreasing the wire diameter from $8.7 L_{\mathrm{ex}}$ to $2.2 L_{\mathrm{ex}}$ the coercive field increased by 40\%.  
In the cylinders nucleation of a reversed domain starts at the corners at the end. Smoothing the edges can improve the coercive field by about 10\%. 

In further simulations we compacted soft magnetic cylinders into a bulk-like arrangement. 
Misalignment and magnetostatic interactions cause a spread of $0.1 M_{\mathrm{s}}$ in the switching fields of the rods. 
Comparing the volume averaged hysteresis loops computed for isolated rods and the hysteresis loop computed for interacting rods, we conclude that magnetostatic interactions reduce the coercive field by up to 20\%. 
\end{abstract}


\end{frontmatter}


\section{Introduction}
\label{sec:Introduction}
Rare earth permanent magnets exhibit the highest maximum energy product $(\boldsymbol{B} \cdot \boldsymbol{H})_{\mathrm{max}}$ of all known magnetic materials; 
a combination of high magnetic moment and high coer\-civ\-ity \cite{kirchmayr1996permanent}.
For efficient electrical generators and motors this is crucial; the permanent magnets that are contained in either the rotor or stator must provide as strong magnetic fields as possible, 
without themselves being demagnetized. 
Recently much of the research into permanent magnets has focussed on reducing their dependence on rare earth elements. 
A number of approaches to finding new permanent magnets are being pursued, including new hard magnetic compounds and nanostructuring, 
where a soft phase contributes large magnet\-ization and a hard phase contributes high coercivity \cite{kneller1991, jiang2004, skomski2013predicting}. 

In this paper we assess the limits of shape anisotropy effects on the improvement of coercivity in magnets made of a single soft magnetic phase. 
The coercive field of an ideal magnetic particle is dependent on size and shape, with various modes of reversal such as coherent rotation, curling and nucleation \cite{coey2004magnetism}. 
At small sizes, where internal magnet\-ization is homogeneous, reversal proceeds by coherent rotation and is described by

\begin{equation}
H_{\mathrm{c}}=\frac{2K_{1}}{\mu _{0}M_{s}} + \frac{M_{s}}{2}(1-3N)
\label{equation:Hc}
\end{equation}
where $K_{1}$ is the uniaxial magneto\-crystalline an\-iso\-tropy constant, $\mu_{0}$ is the vaccuum permeability, $M_{S}$ is the saturation mag\-net\-ization and $N$ is the de\-mag\-net\-izing factor parallel with the $c$ axis. 
The coercive field is dependent on the angle $\theta$ between the field and the $c$ axis of the sample, so that an angle-adjusted coercive field $H_{\mathrm{c}}^{\mathrm{*}}$ is given by \cite{stoner1948mechanism}
\begin{equation}
H_{\mathrm{c}}^{*}=\frac{H_{\mathrm{c}}}{(\mathrm{cos}^{2/3}\theta+\mathrm{sin}^{2/3}\theta)^{3/2}} .
\label{equation:angular}
\end{equation}

For a prolate spheroid (also known as an ``ellipsoid of rotation''), which has equal dimensions along two axes, the de\-mag\-net\-izing factors can be calculated following the work of Osborn \cite{Osborn1945}. 

Previous work \cite{Forster2003,Hertel2002,Escrig2008} showed that, above a certain length, 
the reversal of columnar grains was no longer dependent on grain length but on the nucleation of a reversal volume.  
The nucleation field then depends on the diameter only, where below a certain diameter related to the reversal volume size the nucleation field increases as an inverse function of diameter. 
This is a consequence of the spatial confinement of the reversal volume and the increased exchange energy contribution when forming a domain wall.

\section{Method}
The finite element method is used to numerically solve the Landau-Lifschitz-Gilbert (LLG) equation. At each time step we apply a hybrid finite element/boundary element method to compute
the magnetic scalar potential \cite{schrefl2007numerical}. 
Finite element models are created with tetrahedral volume elements and triangular surface elements. 
Prolate spheroids are prepared by defining inner axes $a$ in the $\mathbf{x-y}$ plane and $b$ along $\mathbf{z}$, with $b\geq a$ defining the long axis (Fig. \ref{fig:spheroidimages}b). 
The dimensions are modified to change the aspect ratio, keeping the total volume  $V=\frac{4}{3}\pi a^{2}b$ constant at $1824 L_{\mathrm{ex}}^{3}$, which corresponds to a sphere with $a=b=4.35 L_{\mathrm{ex}}$ and an aspect ratio of 1.0. 
$L_{\mathrm{ex}}$ is the exchange length of the material. 
In order to assess the effects of shape alone we reduce the magnetocrystalline anisotropy to zero. 

Single phase rods are modelled as regular cylinders with diameter $D$ and length $L$. For comparison, rounded rods are created where the total length remains the same but both ends are rounded spherically. 
For these rods a moderate magneto-crystalline anisotropy is used. 
Reversal loops are calculated by applying a slowly-increasing external field opposing the initial magnetization, where the ramp speed is much slower than the Larmor precession, 
so that dynamic effects may be safely ignored \cite{serpico2007nonlinear}. 
A small field angle of $2^{\circ}$ is used. 
The nucleation field $H_{\mathrm{nuc}}$ is defined as the field strength required to reduce $M_{\mathrm{z}}$ to 0.9$M_{\mathrm{s}}$. 
The coercive field $H_{\mathrm{c}}$ is defined as the field required to reduce $M_{\mathrm{z}}$ to zero. 
For all simulations the $c$ axis is oriented parallel to the cartesian $\mathbf{z}$ axis, except for particle ensembles where the $c$ axis follows the rotation of the geometry, staying with the shape-defined long axis. 

\section{Results \& Discussion}

\begin{figure}[htp]
\includegraphics[width=0.95\columnwidth]{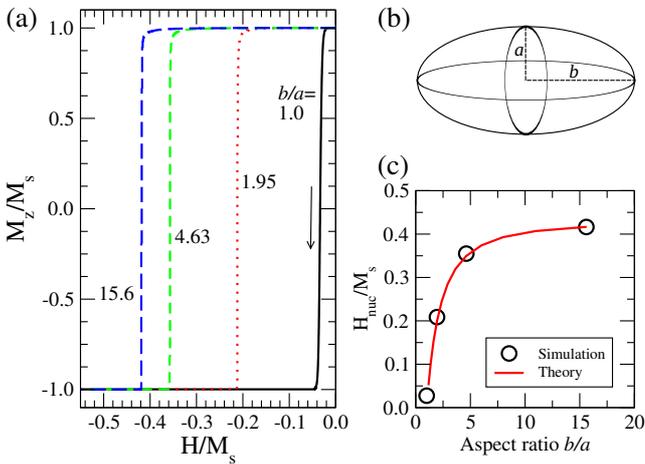}
\caption{
(a) Computed reversal curves for prolate spheroids with zero magnetocrystalline anisotropy and varying aspect ratio between 1.0 and 15.6. 
(b) Schematic of the spheroid geometry, with $a$ being the radius in the $xy$ plane and $b$ along the long $z$ axis. 
(c) The respective nucleation fields for different aspect ratios, with the theoretical plot for coherent reversal. 
}
\label{fig:spheroidloops}
\end{figure}

\begin{figure}[htp]
\begin{center}
\includegraphics[width=0.6\columnwidth]{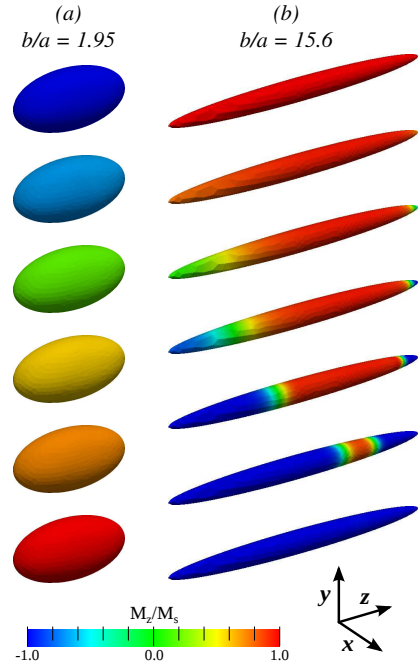} 
\end{center}
\caption{Visualizations of the computed magnet\-ization data during reversal of soft prolate spheroids for aspect ratios (a) 1.95 and (b) 15.6, showing the different reversal modes. }
\label{fig:spheroidimages}
\end{figure}

Numerically calculated hysteresis loops for prolate spheroids with identical volumes but aspect ratios ranging from 1.0 up to 15.6 are presented in Fig. \ref{fig:spheroidloops}a. 
The theoretical values of $H_{\mathrm{c}}$ are calculated with Eq. (\ref{equation:Hc}) and adjusted for the $2^{\circ}$ field angle with Eq. (\ref{equation:angular}). 
The curling mode of reversal is not considered, and images of the changing magnetization configurations during reversal 
for aspect ratios of 1.95 (Fig. \ref{fig:spheroidimages}a) and 15.6 (Fig. \ref{fig:spheroidimages}b) show that curling is not present in this size regime. 

$H_{\mathrm{nuc}}$ approaches zero for an aspect ratio of 1.0, which corresponds to the theoretical coherent mode anisotropy field $H_{\mathrm{A}}=2K/M_{\mathrm{s}}=0$, where for spheres with $N=\frac{1}{3}$ \cite{kronmuller2007handbook}. 
For larger aspect ratios reversal begins with nucleation of a reversal domain at the ends of the wire, followed by domain wall propagation until the whole wire is switched. 
The value of $H_{\mathrm{nuc}}$ converges at a value of $b/a=5$, consistent with previous investigations \cite{Forster2003,Hertel2002,Escrig2008}. 
For an aspect ratio of 15.6 the nucleation field is 0.42$M_{\mathrm{s}}$. 
 
The simulation results match closely with the theoretical model, even at large aspect ratios where reversal begins by nucleation. 
This is consistent with the idea of nucleation of a reversal volume at the ends of the sample, 
since the reversal volume forms through localized rotation from the fully saturated sample (Fig. \ref{fig:spheroidimages}b),  
therefore the Stoner-Wohlfarth model remains valid for predicting the nucleation fields. 
This kind of behaviour has been observed before in the switching of patterned elements in recording media, 
where non-unform switching is observed but the switching field follows the Stoner-Wohlfarh angle dependence \cite{Dittrich2005}.

\begin{figure}[htp,floatfix]
\includegraphics[width=0.75\columnwidth]{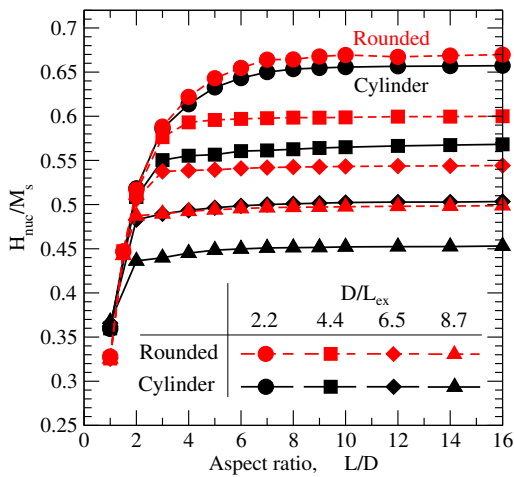}
\caption{Computed nucleation fields for cylindrical (black markers) and rounded (red markers) rods of varying diameter $D$, as a function of wire aspect ratio $L/D$. }
\label{fig:AspectBoth}
\end{figure}

Similar results are presented for cylindrical and rounded rods (Fig.\ref{fig:AspectBoth}), where for a discrete set of rod diameters $D$ the length $L$ is modified. 
The nucleation field $H_{\mathrm{nuc}}$ is plotted as a function of $L/D$, where each symbol represents a different diameter $D$. 
$H_{\mathrm{nuc}}$ values reach a plateau when the aspect ratio is greater than 5. 
In this regime switching occurs by nucleation of a reversed domain and wall motion and the nucleation field increases with decreasing D. 
For each $D$, the coercive field at the plateau is higher for rounded rods. 
The smoothness of the ends reduces surface charges and the resulting lower magneto\-static energy stabilises against nucleation. 
This difference becomes larger as the rod diameter increases, since for smaller diameters the stiffness caused by the exchange interaction limits the inhomogeneity of the magnetization at the sample edges. 
For aspect ratios smaller than 2 we approach the coherent rotation or curling regime. For the rounded ends with an aspect ratio $L/D=1.0$ (a sphere), $H_{\mathrm{nuc}}=0.33M_{\mathrm{s}}$ for all $D$, 
which corresponds to the theoretical Stoner-Wohlfarth prediction for a small sphere of $H_{\mathrm{A}}=0.37 M_{\mathrm{s}}$ (when adjusted for the small field angle) \cite{kronmuller2007handbook}. 
The cylindrical wire with aspect ratio 1.0 has a slightly higher nucleation field, due to the higher shape anisotropy contribution along the parallel axis for cylindrical geometries with respect to spheres \cite{kronmuller2003micromagnetism, chen1991demagnetizing}. 
In this case all of the rounded particles are below the theoretical coherence radius for a sphere $R_{\mathrm{coh}}^{\mathrm{sphere}}=5.099 L_{\mathrm{ex}}$, and all but the largest cylinder are below the cylindrical coherence radius $R_{\mathrm{coh}}^{\mathrm{cylinder}}=3.655 L_{\mathrm{ex}}$, meaning that reversal is coherent \cite{skomski2006nanomagnetic}. 
For larger diameters we expect the nucleation field to reduce, as curling becomes more significant \cite{coey2004magnetism}. 

\begin{figure}[htp]
\includegraphics[width=0.9\columnwidth]{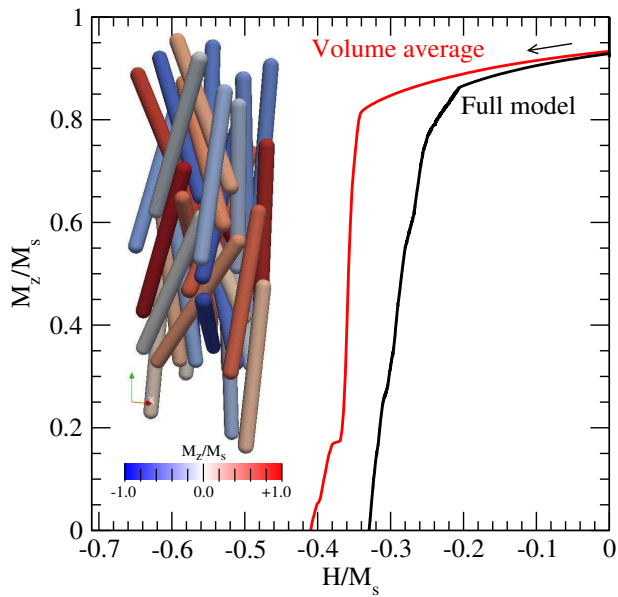}
\caption{The computed reversal curve for the full model rod arrangement shows a 20\% reduction in coercivity compared to the isolated volume-averaged results, due to magnetostatic interactions. Inset: visualization of the packed rods model. }
\label{fig:mikado}
\end{figure}

Exchange-decoupled particle ensembles were created by packing thirty eight of the rounded rods with dimensions $D=8.7 L_{\mathrm{ex}}$ and $L=87.1 L_{\mathrm{ex}}$ into a bounding box of size $65.3 \times 65.3 \times 217.7$ $L_{\mathrm{ex}}^{3}$, 
using the open-source YADE framework (Fig.\ref{fig:mikado}b) \cite{yade:manual}.
Using this method we were able to achieve volume packing densities of between 0.20 and 0.29. 
With this variation, a spread in the coercive field calculated from the hysteresis loops of around $0.1 M_{\mathrm{s}}$ was found. 
This corresponds to a change of up to 30\% and can be attributed to the angular dependence of coercivity as well as the changing distances, orientations and overlap between the individual rods and the accompanying magnetostatic interactions. 

In order to seperate the effect of magnetostatic interactions on the reversal of rod esembles we perform the following experiment: 
Fig. \ref{fig:mikado} compares two computed reversals of the same rod arrangement, with and without magnetostatic interactions. 
The rods have an average tilt angle of $21.8^{\circ}$ and using the same arrangement in both cases excludes the influence of angular distribution. 
First, the full model is simulated at once to include the magnetostatic interactions between the rods. 
Second, the reversal curves for individual rods are computed separately, then the combined volume-average reversal curve is calculated. 
Exchange interactions are not included in order to isolate the influence of magnetostatic interactions. 
The full model reversal exhibits a $0.08 M_{\mathrm{s}}$ reduction in coercivity; a 20\% reduction due to magnetostatic interactions alone. 

We expect that using a larger number of rods would better reproduce the behaviour of bulk magnets and smooth the features of the reversal curve, 
however in this study finite element model sizes were limited to around 38 rods by the available computing resources. 



\section{Conclusions}
By changing the shape of soft magnetic particles from spheres to long wires we were able to increase the nucleation field by a factor of 15. 
Above an aspect ratio of five any improvements plateau. 
The maximum coercive field reaches $0.65 M_{\mathrm{s}}$ for an aspect ratio of 10 and a diameter of the cylinder of $D=2.2 L_{\mathrm{ex}}$. 
Similar results were obtained for cylindrical rods, where smoothing the ends was found to improve the coercivity by reducing surface charges. 
Compacted rod arrangements show a variation in coercivity of around 30\% since different packing arrangements result in different angular distributions and different amounts of magnetostatic interaction between the rods. 
Magnetostatic interactions between the rods were shown to reduce coercivity by around 20\%.





\bibliographystyle{elsarticle-num}
\bibliography{denver.bib}







\end{document}